\documentclass[%
reprint,
 amsmath,amssymb,
 aps,pre,
]{revtex4-1}

\usepackage{graphicx}% Include figure files
\usepackage{dcolumn}% Align table columns on decimal point
\usepackage{bm}% bold math
\usepackage{xcolor}
\newcommand{\st}{\mathrm{st}}
\newcommand{\D}{d}
\newcommand{\bound}{\mathrm{bound}}
\newcommand{\ddev}{\mathrm{dev}}
\newcommand{\hyp}{\mathrm{h}}
\newcommand{\CV}{\mathrm{CV}^2}

\begin{document}

\preprint{APS/123-QED}

\title{Hyperaccurate currents in stochastic thermodynamics}% Force line breaks with \\

\author{Daniel Maria Busiello}
\affiliation{Ecole Polytechnique F\'ed\'erale de Lausanne (EPFL), Institute of Physics Laboratory of Statistical Biophysics, 1015 Lausanne, Switzerland.}%Lines break automatically or can be forced with \\
\author{Simone Pigolotti}%
 \email{simone.pigolotti@oist.jp}
\affiliation{%
Biological Complexity Unit, Okinawa Institute of Science and Technology Graduate University, Onna, Okinawa 904-0495, Japan.
}%

\date{\today}

\begin{abstract}
Thermodynamic observables of mesoscopic systems can be expressed as integrated empirical currents. Their fluctuations are bound by thermodynamic uncertainty relations. We introduce the hyperaccurate current as the integrated empirical current with the least fluctuations in a given non-equilibrium system. For steady-state systems described by overdamped Langevin equations, we derive an equation for the hyperaccurate current by means of a variational principle. We show that the hyperaccurate current coincides with the entropy production if and only if the latter saturates the thermodynamic uncertainty relation, and it can be substantially more precise otherwise. The hyperaccurate current can be used to improve estimates of entropy production from experimental data.
\end{abstract}

\maketitle

%\tableofcontents

Stochastic thermodynamics is a theory describing the non-equilibrium behavior of mesoscopic physical systems, from colloidal particles \cite{seifert2012stochastic,blickle2006thermodynamics,martinez2017colloidal} to molecular motors \cite{julicher1997modeling,pietzonka2016universal,busiello2018similarities}. In these systems, thermodynamic observables are stochastic quantities. A vast class of these observables can be expressed as linear functionals of the increments of a stochastic trajectory. Such observables are called {\em integrated empirical currents}. For continuous systems whose state is specified by a vector $\vec{x}$, an integrated empirical current $R(t)$ (from now on simply ``current") evolves according to the dynamics \cite{chetrite2015nonequilibrium}
\begin{equation}\label{eq:iec}
\frac{\D R}{\D t} = \vec{c}\circ \frac{\D\vec{x}}{\D t}
\end{equation}
where $\vec{c}=\vec{c}(\vec{x})$ is a vector field that determines the current, and $\circ$ indicates the Stratonovich prescription. The total entropy production at steady state and the heat released into a thermal reservoir are examples of thermodynamic observables that can be expressed as currents.

It has been recently observed that, at steady state, all currents must satisfy the so-called {\em thermodynamic uncertainty relation} \cite{barato2015thermodynamic,pietzonka2016universal,gingrich2016dissipation}
\begin{equation}
\frac{\sigma^2_R}{\langle R \rangle^2} \geq \frac{2}{\langle S \rangle}.
\label{eq:bound}
\end{equation}
The left-hand side of Eq.~\eqref{eq:bound} is the coefficient of variation squared ($\CV$) of an arbitrary current $R$, observed at steady state during a time $t$. In the right-hand side, $\langle S\rangle$ is the total entropy produced on average in the same time interval. Equation~\eqref{eq:bound} was originally demonstrated for discrete-state systems described by master equations, first in the long-time limit \cite{gingrich2016dissipation} and later for finite times \cite{horowitz2017proof}. Continuous-state systems described by Langevin equations also satisfy the same bound \cite{dechant2018current}. Interestingly, the bound of Eq.~\eqref{eq:bound} does not hold for discrete-time processes \cite{shiraishi2017finite} and looser bounds have been derived for this case \cite{proesmans2017discrete,chiuchiu2018mapping}. Thermodynamic uncertainty relations have been generalized to periodically driven systems out of steady state \cite{dechant2018multidimensional,koyuk2018generalization} and to observables other than  currents \cite{hasegawa2019fluctuation}. These results have been recently unified with a geometrical interpretation in the space of observables \cite{falasco2019unifying}.

Conceptually, the importance of Eq.~\eqref{eq:bound} is that it sets a universal minimum amount of dissipation necessary to achieve currents of a given precision. Equation ~\eqref{eq:bound} is also of more practical interest: by seeking for currents approaching the bound, one can estimate the entropy production in a more accurate way than with other methods \cite{li2019quantifying}. To this aim, it is important to know which current $R$ approximates the bound best and how close to saturation it is. It was shown that the only current that can saturate the bound is the entropy production itself \cite{hasegawa2019information}. However, it is still unclear what happens when the entropy production does not saturate the bound.

In this Rapid Communication, we introduce the {\em hyperaccurate current} as the current with the lowest $\CV$ in a given stochastic system. For continuous systems described by a set of overdamped Langevin equations, we derive the Euler-Lagrange equations that must be satisfied by the hyperaccurate current, and solve them in concrete examples.

We consider mesoscopic physical systems that can be described by $N$ slow degrees of freedom $\vec{x}=\vec{x}(t)=x_1(t),x_2(t),\dots, x_N(t)$. Such degrees of freedom evolve according to a set of overdamped Langevin equations
\begin{equation}\label{eq:langevin}
\frac{\D}{\D t} \vec{x} = \hat{\mu} \cdot \vec{F} + \vec{\nabla} \cdot \hat{D} + \sqrt{2} \hat{\sigma} \cdot \vec{\xi}
\end{equation}
where $\vec{\xi}=\vec{\xi}(t)=\xi_1(t),\dots,\xi_N(t)$ is a Gaussian white noise with mean $\langle \xi_i(t) \rangle = 0$ and autocorrelation $\langle \xi_i(t) \xi_j(t') \rangle = \delta(t-t') \delta_{ij}$. Here the noise is interpreted in the Ito sense. The symmetric matrix $\hat{\mu}=\hat{\mu}(\vec{x})$ is the motility tensor and the vector $\vec{F}=\vec{F}(\vec{x})$ is the force acting on the system. The matrix $\hat{\sigma}=\hat{\sigma}(\vec{x})$ is related to the symmetric diffusion matrix $\hat{D}=\hat{D}(\vec{x})$ by the relation $\hat{\sigma}^T\hat{\sigma}=\hat{D}$. We assume the Einstein relation $\hat{D}=k_B T\hat{\mu}$ to hold, where $k_B$ is the Boltzmann constant and $T$ the temperature. We further assume that the matrices $\hat{\sigma}$, $\hat{D}$, and $\hat{\mu}$ are non-degenerate. We associate to Eqs. \eqref{eq:langevin} the Fokker-Planck equation
\begin{equation}\label{eq:FP}
\partial_t P(\vec{x};t)=\vec{\nabla} \cdot \left[- \hat{\mu} \cdot \vec{F}P(\vec{x};t)+\hat{D} \cdot \vec{\nabla} P(\vec{x};t)\right].
\end{equation}
We call $P^\st=P^\st(\vec{x})$ the stationary solution of Eq. \eqref{eq:FP}, $P(\vec{x};t|\vec{y};t')$ the propagator, $\vec{J}=\vec{J}(\vec{x},t)=\hat{\mu} \cdot \vec{F}P(\vec{x},t)-\hat{D} \cdot \vec{\nabla} P(\vec{x},t)$ the flux, and $\vec{J}^{\st} = \vec{J}^{\st}(\vec{x})=\hat{\mu} \cdot \vec{F}P^\st(\vec{x})-\hat{D} \cdot \vec{\nabla} P^\st(\vec{x})$ the stationary flux.
We substitute Eqs. \eqref{eq:langevin} and \eqref{eq:FP} into Eq. \eqref{eq:iec}, finding an explicit evolution equation for a generic current
\begin{eqnarray}\label{eq:currentito}
\frac{\D R}{\D t} &=& \frac{\vec{c} \cdot \vec{J} + \vec{\nabla} \cdot (\hat{D} \cdot \vec{c}~ P)}{P} + \sqrt{2} \vec{c} \cdot \hat{\sigma} \cdot \vec{\xi} .
\end{eqnarray}

Equation \eqref{eq:currentito} is interpreted in the Ito sense.
Important examples of currents are the heat released in the thermal bath $Q$, with $\vec{c}=\vec{F}$ \cite{sekimoto2010stochastic}, 
and the total entropy production $S$ at steady state, with $\vec{c}=\hat{D}^{-1} \cdot \vec{J}^{\st}/P^\st$.
Substituting this latter choice into Eq. \eqref{eq:currentito} directly yields the evolution equation for the entropy production derived in \cite{pigolotti2017generic}.

We consider the evolution a current at steady state and use Eq.~\eqref{eq:currentito} to derive the uncertainty bound of Eq.~\eqref{eq:bound} in a straightforward way. We introduce the {\em bound term}
\begin{equation}
\frac{\D R_\bound}{\D t}= \sqrt{2} \frac{\langle R \rangle}{\langle S\rangle} \frac{\vec{J}^\st}{P^\st} \cdot (\hat{\sigma}^T)^{-1} \cdot \vec{\xi}.
\label{eq:bound2}
\end{equation}
The bound term is defined so that its variance over the mean of the current squared saturates the uncertainty bound of Eq.~\eqref{eq:bound}, i.e.,
\begin{equation}
\frac{\sigma^2_{R_\bound}}{\langle R \rangle^2} = \frac{2}{\langle S\rangle}.
\end{equation}
We now decompose an arbitrary  current $R(t)$ into the sum of the bound term and a deviation term
\begin{equation}
R_{\rm dev}(t) = R(t) - R_\bound(t).
\label{eq:defdev}
\end{equation}
In terms of this decomposition, the left-hand side of the uncertainty bound reads
\begin{equation}
\frac{\sigma^2_{R}}{\langle R \rangle^2} = \frac{\sigma^2_{R_{\rm bound}}}{\langle R \rangle^2} + \frac{\sigma^2_{R_\ddev}}{\langle R \rangle^2} + 2 \frac{\sigma^2_{R_{\rm bound},R_{\rm dev}}}{\langle R \rangle^2}.
\end{equation}
An explicit computation shows that the covariance $\sigma^2_{R_\bound,R_\ddev}$ always vanishes, see SI. This implies
\begin{equation}
\frac{\sigma^2_{R}}{\langle R \rangle^2} = \frac{2}{\langle S\rangle} + \frac{\sigma^2_{R_\ddev}}{\langle R \rangle^2} \geq \frac{2}{\langle S\rangle}.\label{eq:dev}
\end{equation}
Equation \eqref{eq:dev} means that the variance of $R_\ddev$ is responsible for the deviation from the bound.

This calculation constitutes a short and direct demonstration of the thermodynamic uncertainty relation for a system governed by Langevin equations \cite{dechant2018current}. An advantage of this approach is to provide an explicit expression for the deviation from the bound. In particular, a current $R$ saturates the
uncertainty bound only when $\sigma^2_{R_\ddev}=0$. A necessary
condition for this to hold is that the noise amplitude of $R_\ddev$ must vanish. 
Imposing this condition by means of Eqs.~\eqref{eq:currentito}, \eqref{eq:bound2}, and \eqref{eq:defdev} yields
\begin{equation}
\sqrt{2}\left( \vec{c} - \frac{\vec{J}^{\st}}{P^{\st}} \cdot \hat{D}^{-1} \frac{\langle R \rangle}{\langle S\rangle} \right) \cdot \hat{\sigma} = 0 \;\;\;\ \Leftrightarrow \;\;\;\ \vec{c} \propto \frac{\vec{J}^{\st}}{P^{\st}} \cdot \hat{D}^{-1}.
\label{cstar}
\end{equation}
When $\vec{c}$ satisfies the condition in Eq.~\eqref{cstar}, then
$R \propto S$. This means that only the entropy production, or a current proportional to it, can saturate the uncertainty bound \cite{hasegawa2019information}.  As a corollary, if the entropy production does not saturate the bound, the bound can not be saturated by any current.

To understand such cases, we define the {\em hyperaccurate current} $R_\hyp$ as the current with the minimum $\CV$, among all possible choices of $\vec{c}(\vec{x})$. Since $\sigma^2_R/\langle R\rangle^2= \langle R^2\rangle /\langle R\rangle^2-1$, we seek for the hyperaccurate current by minimizing $\langle R^2\rangle /\langle R\rangle^2$ with respect to the function $\vec{c}(\vec{x})$.

The average value of $R$ reads
\begin{equation}\label{eq:rav}
\langle R\rangle=t\left\langle\frac{\D R}{\D t}\right\rangle= t\int \D \vec{x} ~\vec{c}(\vec{x}) \cdot \vec{J}^{\st}(\vec{x})~ 
\end{equation}
where in the last equality we used Eq.~\eqref{eq:currentito}. Similarly, we express the second moment as $\langle R^2\rangle = \langle [\int_0^t \D t'  (\D R/\D t')]^2  \rangle$. We use these expressions to evaluate the first variation of $\langle R^2\rangle/\langle R\rangle^2$ with respect to $\vec{c}(\vec{x})$ and impose that it must vanish (see \cite{SI}). This procedure results in the Euler-Lagrange equation

\begin{eqnarray}
 &&  \hat{D}^{-1}(\vec{x}) \cdot \vec{J}^{\st}(\vec{x}) \int_0^t \D t' \int_0^{t'} \D t''  \left\langle \frac{\vec{J}^{\st}(\vec{y}) \cdot\vec{c}_\hyp (\vec{y})}{P^\st(\vec{x})P^\st(\vec{y})}\right\rangle_y + \nonumber \\
&+& P^\st(\vec{x}) \vec{\nabla}_{\vec{x}} \left\{ \int_0^t \D t' \int_0^{t'} \D t''  \left\langle ~\frac{\vec{\nabla}_{\vec{y}}\cdot\left[ P^\st(\vec{y})\hat{D}(\vec{y}) \cdot \vec{c}_\hyp (\vec{y})\right]}{P^\st(\vec{x}) P^\st(\vec{y})}  \right\rangle_y \right\} = \nonumber \\
&=& t \hat{D}^{-1}(\vec{x}) \cdot \vec{J}^{\st}(\vec{x}) \frac{\langle R_\hyp^2\rangle}{2\langle R_\hyp \rangle}-t P^\st(\vec{x}) \vec{c}_\hyp (\vec{x})
\label{eq:gforward}
\end{eqnarray}
where $\vec{c}_\hyp (\vec{x})$ is the vector field associated to the hyperaccurate current, and we denoted with $\langle\dots\rangle_y=\int \D\vec{y} P(\vec{x};t|\vec{y};t'')P^\st(\vec{y})$ the average over the initial state. In principle, also the Fano factor $\langle R_\hyp^2\rangle/(2\langle R_\hyp \rangle)$  on the right-hand side of Eq. \eqref{eq:gforward} implicitly depends on $\vec{c}_\hyp(\vec{y})$. However, we can exploit the fact that rescaling $\vec{c}_\hyp(\vec{y})$ by an arbitrary multiplicative factor does not change its $\CV$. The solution of Eq. \eqref{eq:gforward} is therefore defined up to an arbitrary multiplicative constant. From now on, we shall fix this constant by setting $ \sigma^2_{R_\hyp}/2\langle R_\hyp \rangle=1$.

In the long time limit, Eq. \eqref{eq:gforward} reduces to the simpler form
\begin{equation}
\int \D \vec{y} ~\hat{K}(\vec{x},\vec{y})\cdot \vec{c}_\hyp(\vec{y})=\vec{J}^{\st}(\vec{x}) 
\label{eq:kerneleq}
\end{equation}
see SI, where we defined the integral kernel
\begin{gather}
\hat{K}(\vec{x},\vec{y})=\frac{\vec{J}^{\st}(\vec{x})}{P^\st(\vec{x})}\phi(\vec{x},\vec{y})  \vec{J}^{\st}(\vec{y})+P^\st(\vec{x})\hat{D}(\vec{x})\delta(\vec{x}-\vec{y})\nonumber\\
- P^\st(\vec{x})\left[ \hat{D}(\vec{x})\cdot
\vec{\nabla}_{\vec{x}}\right]\cdot \vec{\nabla}_{\vec{y}} \left[\frac{\phi(\vec{x},\vec{y})}{P^\st(\vec{x})}\right]\cdot \hat{D}(\vec{y})P^\st(\vec{y}).
\label{eq:kernel}
\end{gather}
and the function
\begin{equation}
 \phi(\vec{x},\vec{y}) = \int_0^{+\infty} \D t ~ \left[P(\vec{x};t | \vec{y};0) - P^\st(\vec{x})\right].
\end{equation}
If the kernel $\hat{K}(\vec{x},\vec{y})$ can be inverted, then $\vec{c}_\hyp(\vec{x})$ can be expressed as
\begin{equation}
\vec{c}_\hyp(\vec{x}) = \int \D\vec{y} ~\hat{K}^{-1}(\vec{x},\vec{y}) \cdot \vec{J}^{\st}(\vec{y})
\label{eq:explicitsol}
\end{equation}
where $\int \D z ~\hat{K}^{-1}(\vec{x},\vec{z}) \cdot \hat{K}(\vec{z},\vec{y}) = \delta(\vec{x}-\vec{y})$.

We are now in the position to study whether the entropy production can still be hyperaccurate when it does not saturate the bound. To this aim, we assume $R_\hyp \propto S$, i.e., $\vec{c}_\hyp \propto \hat{D}^{-1} \cdot \vec{J}^{\st} / P^{\rm st}$ and substitute this choice into Eq.~\eqref{eq:kerneleq}, obtaining
\begin{gather}
\int \D \vec{y} ~\phi(\vec{x},\vec{y}) \frac{\vec{J}^{\st}(\vec{y}) \cdot \hat{D}^{-1}(\vec{y}) \cdot \vec{J}^{\st}(\vec{y})}{P^{\rm st}(\vec{y})} \propto P^\st(\vec{x}).
\label{eq:RasS}
\end{gather}

We interpret the left hand side of Eq. \eqref{eq:RasS} as the integral operator $\int \D\vec{y}~ \phi(\vec{x},\vec{y})$ acting on the function $g(\vec{y})=\vec{J}^{\st}(\vec{y}) \cdot \hat{D}^{-1}(\vec{y}) \cdot \vec{J}^{\st}(\vec{y})/P^{\rm st}(\vec{y})$. Such integral operator shares the same eigenfunctions of the Fokker-Planck equation \eqref{eq:FP}. In particular, the stationary solution in the right-hand side of Eq. \eqref{eq:RasS} is a  right eigenfunction associated to a non degenerate eigenvalue equal to zero. Therefore, Eq. \eqref{eq:RasS} can be satisfied only if $g(y)\propto P^\st(\vec{y})$, i.e., if the quantity $\vec{J}^{\st} \cdot \hat{D}^{-1} \cdot \vec{J}^{\st} / (P^{\rm st})^2$ is constant. But this is precisely the condition for the entropy production to saturate the uncertainty bound \cite{pigolotti2017generic}. We therefore conclude that, when the entropy production does not saturate the bound, it cannot be identified as the hyperaccurate current.

By definition, the $\CV$ of the hyperaccurate current provides the tightest possible bound on the $\CV$ of a current, the \textit{hyperaccurate bound} $\mathcal{B}_\hyp$. Since we set $\sigma^2_{R_\hyp}/2\langle R_\hyp \rangle=1$, $\mathcal{B}_\hyp$ depends solely on the average of $R_\hyp$
\begin{equation}
\frac{\sigma^2_R}{\langle R \rangle^2} \geq \mathcal{B}_\hyp = \frac{\sigma^2_{R_\hyp}}{\langle {R_\hyp} \rangle^2} = \frac{2}{\langle R_\hyp \rangle}.
\end{equation}
By using Eqs. \eqref{eq:rav} and ~\eqref{eq:explicitsol} to express the average of the hyperaccurate current, we obtain
\begin{equation}
\mathcal{B}_\hyp = \frac{2}{t} \left( \int \D\vec{x}~ \D\vec{y} ~\vec{J}^{\st}(\vec{y}) \cdot \hat{K}^{-1}(\vec{y},\vec{x}) \cdot \vec{J}^{\st}(\vec{x}) \right)^{-1}.
\end{equation}

We now study the hyperaccurate current in two concrete models, where we take $\hat{\mu}=\hat{D}=\hat{I}$ for simplicity, with $\hat{I}$ the identity matrix. Our first example is  a molecular motor in a one-dimensional periodic potential $U(x) = \sin(2 \pi x)$ subject to a constant non-conservative force $f$. The system is described by the Langevin equation
\begin{equation}\label{eq:motor}
\frac{\D x}{\D t} = f  - \frac{\D U(x)}{\D x} + \sqrt{2} \xi.
\end{equation}

In this case, Eq.~\eqref{eq:kernel} is one dimensional. We numerically solve it by discretizing the interval $[0,1]$ with a mesh $\Delta$, so that the  integral in Eq. \eqref{eq:kerneleq} becomes a linear system of equations and the integral kernel in Eq. \eqref{eq:kernel} becomes a matrix. We estimate this matrix by solving the Fokker-Planck equation numerically with the same spatial mesh $\Delta$ (see \cite{SI} for details).

\begin{figure}[th]
    \centering
    \includegraphics[width= \columnwidth]{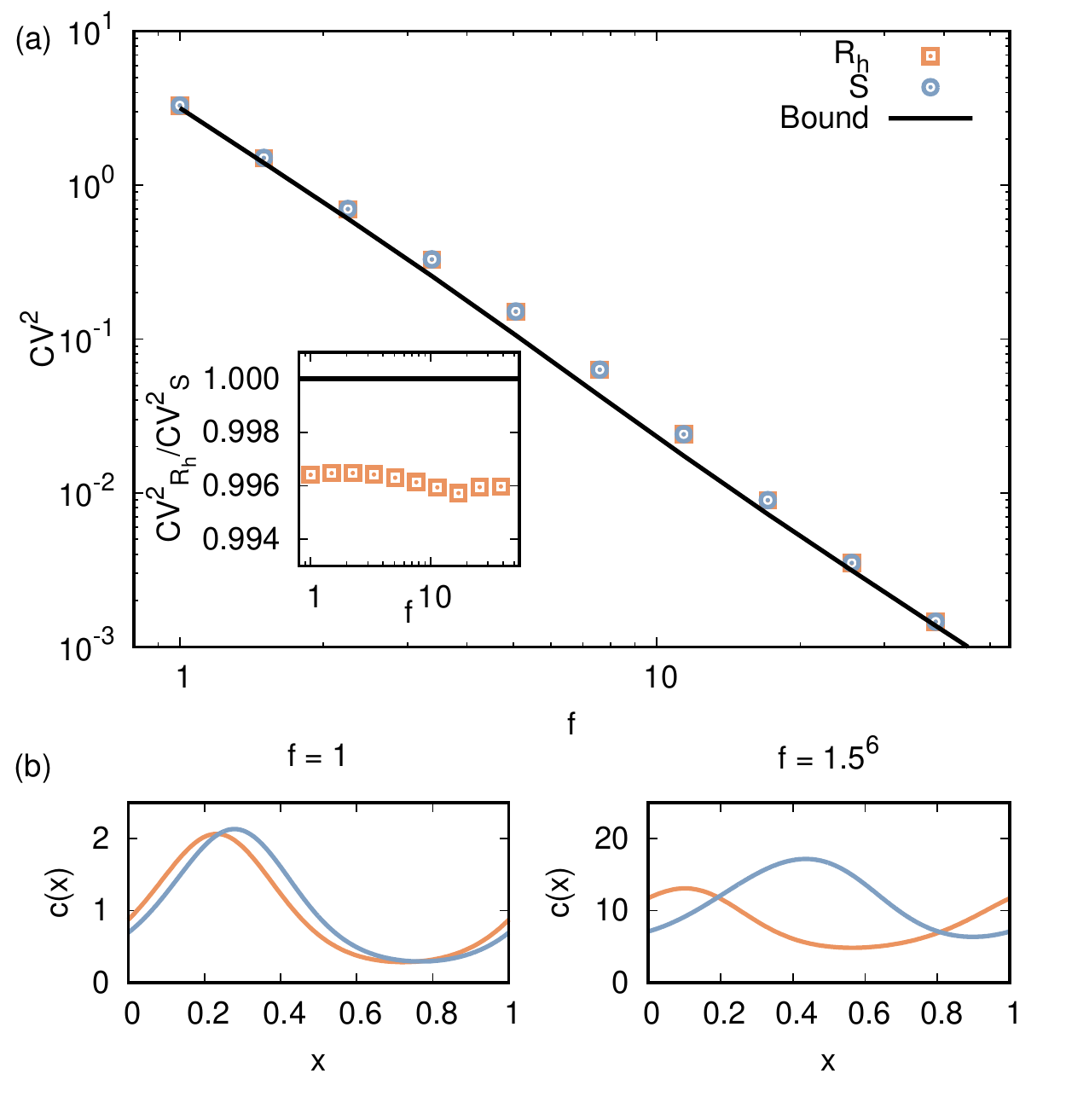}
    \caption{Hyperaccurate current of a molecular motor model,  Eq. \eqref{eq:motor}. (a) $\CV$ of the hyperaccurate current and the entropy production as a function of the force $f$. The continuous line is the uncertainty bound of Eq. \eqref{eq:bound}. Inset: Ratio between the $\CV$ of the hyperaccurate current and that of the entropy production as a function of $f$. (b) Comparison of $c(x)$ for the hyperaccurate current in red (lighter gray) and for the entropy production in blue (darker gray) for two different values of the force $f$, shown in the figures. }
    \label{fig:1D}
\end{figure}

In this model, both $R_\hyp$ and $S$ are quite close to the bound, Fig.~\ref{fig:1D}(a), with appreciable differences only for intermediate values of $f$ (see also \cite{pigolotti2017generic}). The $\CV$ of $R_\hyp$ is lower than that of $S$ as predicted, although their difference is rather small [less than $1\%$ in the range of $f$ we considered; inset of Fig.~\ref{fig:1D}(a)]. Inspecting $c_\hyp(x)$, we find that it is rather similar to the one characterizing the entropy production for low values of the force and substantially different at larger values of the force, Fig. \ref{fig:1D}(b).

As a second example, we consider the two dimensional Langevin dynamics on a torus $[0,1] \times [0,1]$:
\begin{eqnarray}
\frac{\D x_1}{\D t} &=& F(x_2) + \sqrt{2} \xi_1 \nonumber \\
\frac{\D x_2}{\D t} &=& \sqrt{2} \xi_2
\label{2Dsystem}
\end{eqnarray}
with the non-conservative force $F(x_2) = f \cos(2 \pi x_2)$. The stationary probability distribution is homogeneous, $P^\st(x_1,x_2)=1$ and the steady state flux is $\vec{J}^{\st}(x_1,x_2) = F(x_2),0$. 

Since the dynamics is invariant under translations along the $x_1$ axis, then $\vec{c}_\hyp(x_1,x_2)=c_{\hyp,1}(x_1,x_2),c_{\hyp,2}(x_1,x_2)$ cannot depend on $x_1$. Writing Eq.~\eqref{eq:kerneleq} by components, we find that $c_{\hyp,2}(x_2) = 0$ (see SI). Consequently, Eq.~\eqref{eq:kerneleq} reduces to the one-dimensional equation in the unknown $c_{\hyp,1}(x_2)$
\begin{equation}
\int \D y_2 ~ K(x_2,y_2) c_{\hyp,1}(y_2) = f \cos(2 \pi x_2)
\label{eq:sol2D}
\end{equation}
where the kernel is
\begin{equation}
K(x_2,y_2) = f^2 \cos(2 \pi x_2) \phi(x_2,y_2) \cos(2 \pi y_2) + \delta(x_2 - y_2).
\label{kernel2D}
\end{equation}
Since the coordinate $x_2$ evolves according to a simple diffusion process with periodic boundary conditions, the function $\phi(x_2,y_2)$ can be explicitly expressed as
\begin{equation}
\phi(x_2,y_2) = \sum_{n=0}^{+\infty} \frac{1}{2 \pi^2 n^2} \cos[2 \pi n (x_2 - y_2)]
\end{equation}

Expanding the solution $c_{\hyp,1}(x_2)$ in a Fourier basis and substituting into Eq.~\eqref{eq:sol2D}, the Fourier coefficients can be analytically calculated at any order (see \cite{SI}).

\begin{figure}[th]
    \centering
    \includegraphics[width= \columnwidth]{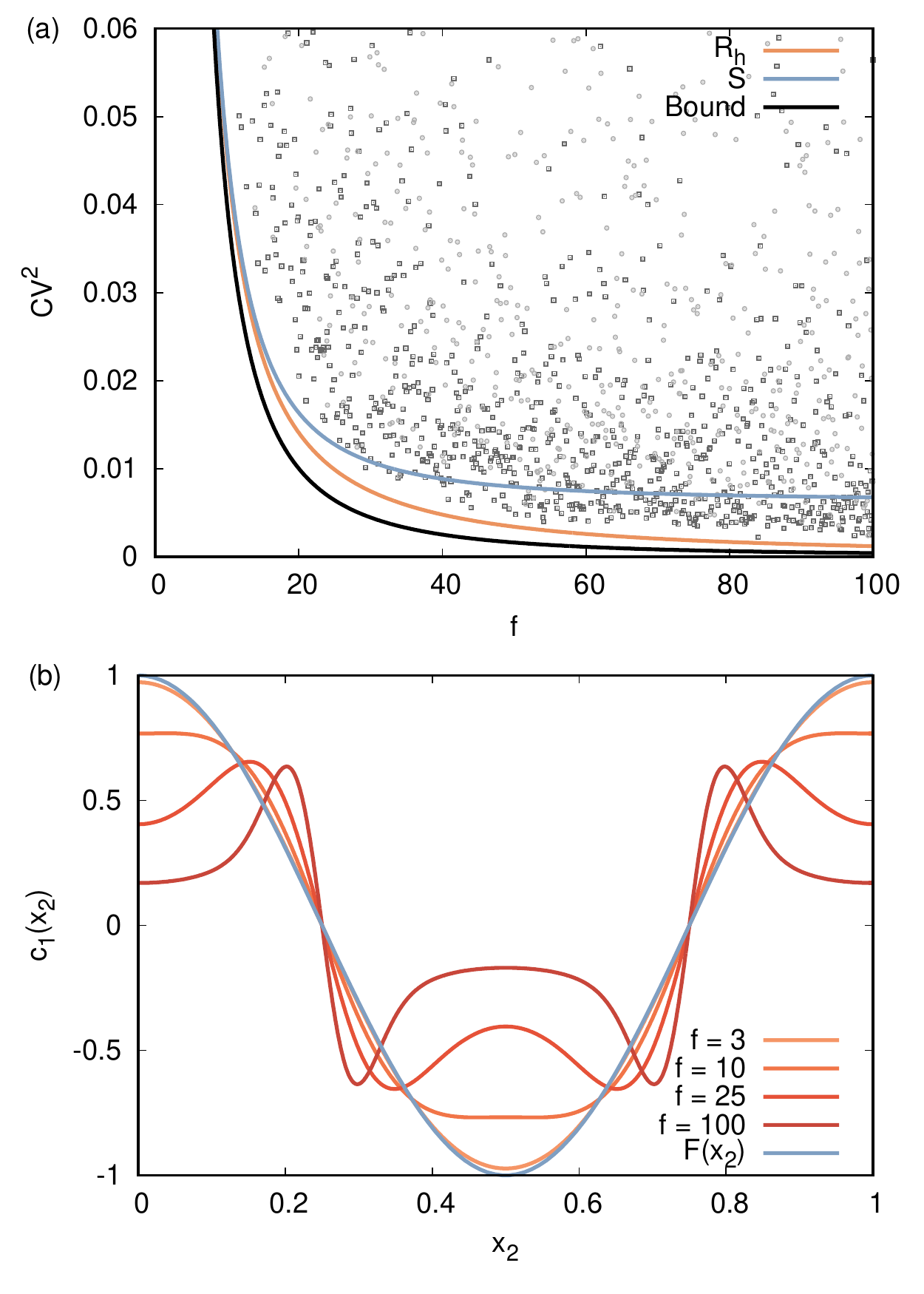}
    \caption{Hyperaccurate current in a two-dimensional model,  Eq. \eqref{2Dsystem}. (a) The black line is the thermodynamic 		uncertainty bound. The blue (top gray) line is the $\CV$ of the hyperaccurate current. The red (middle gray) line is the $\CV$ of the entropy production. All curves are plotted as a function of the nonconservative force $f$. The points represent random currents generated by adding to the coefficients $\kappa_{h,n}$ Gaussian random variables with mean zero and variance equal to $f$ (dark-gray points) and $4f$ (light-gray points). (b) Red (dark-gray) lines represent $c_{h,1}(x_2)$ for different values of $f$. The blue line (dark-gray) represents $c_1(x_2) = F(x_2)$, whose associated current is the entropy production.}
    \label{fig:2D}
\end{figure}

In this case the $\CV$ of the hyperaccurate current is much lower than that of the entropy production far from equilibrium, i.e., when $f \gg 1$ [see Fig.~\ref{fig:2D}(a)]. The hyperaccurate current converges to the entropy production when the system is near equilibrium and the bound tends to be saturated. Farther from equilibrium, the hyperaccurate current is markedly different from the entropy production,  Fig.~\ref{fig:2D}(b).

In this Rapid Communication, we introduced the hyperaccurate current for systems described by overdamped Langevin equations. We have shown with examples that the hyperaccurate current can be substantially more accurate than the entropy production, in cases where the latter significantly departs from the uncertainty bound. By its definition, the hyperaccurate current provides the tightest possible uncertainty bound to the $\CV$ of an arbitrary current. Our theory can be extended to discrete-state or discrete-time systems and possibly employed to study non integrated currents or non stationary dynamics. We leave these investigations for future work.

It is worthwhile discussing how the results presented here can help in estimating entropy production in experiments. Naive estimators of entropy production often require very large sample size and/or observation times to provide accurate results. Reference \cite{li2019quantifying} proposes to use Eq.~\eqref{eq:bound} as a tool to estimate entropy production, or at least bound it. This strategy relies on the fact that, empirically, the $\CV$ of a current is much easier to estimate than the entropy production. One crucial ingredient of this strategy is to identify a current whose $\CV$ is sufficiently close to the bound. Reference \cite{li2019quantifying} tackles this problem by means of a Monte Carlo scheme. This approach is relatively simple to implement, but has the disadvantages of being computationally costly and prone to overfitting, especially in high-dimensional systems. These difficulties are circumvented by the theory developed in this Rapid Communication. One possible strategy is therefore to build an approximate model of the physical system at hand, evaluate its hyperaccurate current using the theory developed in this Rapid Communication, and then measure the $\CV$ of the hyperaccurate current in experiments. To pursue this strategy, it will be key to develop efficient numerical schemes \cite{delves1988computational} to solve the integral equation \eqref{eq:kerneleq} and therefore compute the hyperaccurate current in systems more complex than the simple examples considered in this Rapid Communication. The results of Fig. 2(a) show that, even perturbing the hyperaccurate current, one can obtain currents that are substantially more accurate than the entropy production. This supports the idea that the hyperaccurate current computed in an approximate model of a physical system can be sufficiently close to the bound to provide a reliable estimate of entropy production, if measured in an experiment.
\\

We acknowledge D. Chiuchi\'{u}, E. Fried, F. J\"{u}licher, A. Maritan, I. Neri,  L. Peliti, and \'{E} Rold\'{a}n for many discussions.

\providecommand{\noopsort}[1]{}\providecommand{\singleletter}[1]{#1}%

\clearpage

\begin{widetext}

\section*{Supplementary Information}

In this document, we provide additional calculations and mathematical details complementing the manuscript ``Hyperaccurate currents in stochastic thermodynamics" (from now on ``Main Text"). The document is organized as follows. In Section~\ref{I}, we show how to evaluate averages of stochastic integrals that we often use in the following. In Section~\ref{II} we show that $\sigma^2_{R_\bound,R_\ddev}=0$. In Section~\ref{III} we derive the Euler-Lagrange equation for the vector field $\vec{c}_\hyp(\vec{x})$ defining the hyperaccurate current. The long-time limit is presented in Section~\ref{IV}.  In Section~\ref{V} we discretize the integral kernel of the Euler-Lagrange equation for solving it numerically on a one-dimensional grid. Finally, in Section~\ref{VI} we analytically compute $\vec{c}_\hyp(\vec{x})$  for the two-dimensional model presented in the Main Text.

\section{Evaluating two-time averages with the Doob transform}\label{sec:doob}
\label{I}

In this Section, we show how to evaluate averages of stochastic integrals of the form

\begin{equation}\label{eq:doobtrick}
I=\left\langle \int~\D t' \int \D t''~\vec{f}(\vec{x}(t'),t',\vec{y}(t''),t'') \cdot\vec{\xi}(t'') \right \rangle
\end{equation}
where $\vec{f}(\vec{x}(t'),t',\vec{y}(t''),t'')$ is an arbitrary function of
an Ito process at two different times (see also \cite{pigolotti2017generic}).

For $t''\ge t'$, the average in Eq. \eqref{eq:doobtrick} always vanishes due to the non-anticipating properties of the Wiener process in the Ito calculus. The case $t''<t'$ requires more care. To evaluate the average in this case, we introduce the Doob transform of the process.

Doob transform maps a stochastic process conditioned on a future event to an unconditioned stochastic process with an additional drift term. In our case, we consider the Langevin equation:
\begin{equation}
\frac{\D}{\D t} \vec{x} = \hat{\mu} \cdot \vec{F} + \vec{\nabla} \cdot \hat{D} + \sqrt{2} \hat{\sigma} \cdot \vec{\xi}
\label{Langevin}
\end{equation}
and impose a future condition $\vec{x}(t') = \vec{x}'$, with $t' > t$.  Doob showed that the ensemble of trajectories $\{\vec{x}\}$ of Eq. ,\eqref{Langevin} conditioned on the future event $\vec{x}(t') = \vec{x}'$ is equal to the unconditioned ensemble of trajectories $\{\vec{z}\}$ generated by the Langevin equation
\begin{equation}
\frac{\D}{\D t} \vec{z} = \hat{\mu} \cdot \vec{F} + \vec{\nabla} \cdot \hat{D} + \sqrt{2} \hat{\sigma} \cdot \vec{\eta}(t) + 2 \hat{D} \cdot \vec{\nabla}_{\vec{y}} \log P(\vec{x}; t'|\vec{y}; t'')
\label{LangevinDoob}
\end{equation}
where $\vec{\eta}(t'')$ is unbiased white noise. Comparing Eq. \eqref{Langevin} and Eq. \eqref{LangevinDoob}, we obtain that the conditioned averages can be transformed into unconditioned ones by substituting the noise term
\begin{equation}
\vec{\xi}(t) = \vec{\eta}(t) + \sqrt{2} \hat{\sigma}^T \vec{\nabla}_{\vec{y}} \log P(\vec{x}; t'|\vec{y}; t).
\end{equation}
The second term in the right hand side represents an additional drift. Substituting
this expression in Eq. \eqref{eq:doobtrick} and using that the average of any function multiplied by $\vec{\eta}$ vanishes, we obtain

\begin{equation}\label{eq:doobtrick2}
I=\sqrt{2}\left\langle \int~\D t' \int \D t''~ \vec{f}(\vec{x}(t'),t',\vec{y}(t''),t'')\cdot  \hat{\sigma}^T \cdot\vec{\nabla}_{\vec{y}} \log P(\vec{x}; t'|\vec{y}; t'')\cdot \right \rangle
\end{equation}

\section{Proof that $\sigma^2_{R_\bound,R_\ddev}=0$}
\label{II}

We now demonstrate that the covariance term $\sigma^2_{R_\bound,R_\ddev}=\langle R_\bound R_\ddev\rangle-\langle R_\bound\rangle\langle R_\ddev\rangle$ vanishes, where $R_\bound$ and $R_\ddev$ are defined in Eqs. (6) and (8) of the Main Text.  Since $\langle R_\bound\rangle=0$, we express the covariance as
\begin{eqnarray}\label{eq:covar}
\sigma^2_{R_\bound,R_\ddev}&=&\langle R_\bound~(R-R_\bound) \rangle=\nonumber\\
&=& - \sigma^2_{R_{\rm bound}} + 2 \frac{\langle R \rangle}{\langle \Sigma \rangle} \left\langle \frac{\vec{J}^{\st}}{P^{\st}} \cdot (\hat{\sigma}^T)^{-1} \hat{\sigma}^T \cdot \vec{c} \right\rangle + \nonumber \\
&+& 2 \frac{\langle R \rangle}{\langle \Sigma \rangle} \int_0^t \D t' \int_0^{t'} \D t'' \int \D \vec{x}~ \D \vec{y}~ \frac{\psi(\vec{y})}{P^{\st}(\vec{y})} \frac{\vec{J}^{\st}(\vec{y})}{P^{\st}(\vec{y})} \cdot [\vec{\nabla}_{\vec{y}} P(\vec{x} t'|\vec{y} t'') ]P^{\st}(\vec{y})
\end{eqnarray}
where in the last term we expressed the average over the noise $\xi(t'')$ conditioned at future time using the Doob transform, see Section \ref{sec:doob}. The quantity $\psi$ is defined as
\begin{equation}\label{eq:phi}
\psi(\vec{x})=\vec{c} (\vec{x})\cdot \vec{J}^{\st}(\vec{x}) + \vec{\nabla}_{\vec{x}} \cdot [\hat{D}(\vec{x}) \cdot \vec{c}(\vec{x}) P^{\st}(\vec{x})].
\end{equation}
We now rewrite the last expression in Eq. \eqref{eq:covar} using that $\langle \vec{J}^\st\cdot\vec{c}/P^\st\rangle = \int \D \vec{x}~ \vec{J}^\st(\vec{x})\cdot\vec{c}(\vec{x})=\langle R\rangle$ (see Eq. (12) in the Main Text) and integrating by parts over $\vec{y}$ the last term. We obtain 
\begin{equation}
\sigma^2_{R_\bound,R_\ddev} = - 2 \frac{\langle R \rangle^2}{\langle \Sigma \rangle} + 2 \frac{\langle R \rangle^2}{\langle \Sigma \rangle} + 2 \frac{\langle R \rangle}{\langle \Sigma \rangle} \int_0^t \D t' \int_0^t \D t'' \int ~\D \vec{x} ~\D \vec{y}~ \frac{\psi(\vec{x})}{P^{\st}(\vec{x})} P(\vec{x}; t'|\vec{y}; t'') \vec{\nabla}_{\vec{y}} \cdot \vec{J}^{\st}(\vec{y}) = 0
\end{equation}
since $\vec{\nabla}_{\vec{y}}\cdot\vec{J}^{\st}(\vec{y}) = - \partial_t
P^{\st}(\vec{x}) = 0$ at steady state. Here and in the following, when integrating by parts  we assume that the boundary term always vanish due to appropriate boundary conditions on $P(\vec{x})$. 
This result directly leads to Eq.~(10) in the Main Text.
\section{Derivation of the Euler-Lagrange equations}
\label{III}
In this Section we derive the Euler-Lagrange equations for the hyperaccurate current. We want to minimize the quantity $\langle R^2\rangle/\langle R\rangle^2$. Our first step is to derive tractable expressions for the first two moments of the current. The average reads
\begin{equation}
\langle R\rangle=t\left\langle \frac{\vec{c} \cdot \vec{J}^{\st}}{P^{\st}}\right\rangle
\end{equation}
The second moment can be expressed as
\begin{eqnarray}\label{eq:tominimize}
\langle R^2\rangle &=& 2  t\langle \vec{c} \cdot \hat{D} \cdot \vec{c} \rangle + 2 \int_0^t \D t' \int_{0}^{t'} \D t''\left[\left\langle \frac{\psi(\vec{x})}{P^{\st}(\vec{x})} \frac{\psi(\vec{y})}{P^{\st}(\vec{y})} \right\rangle + \sqrt{2}\left\langle \frac{\psi(\vec{x})}{P^{\st}(\vec{x})} \vec{c}(\vec{y}) \cdot \hat{\sigma}^{T}(\vec{y})\cdot \vec{\xi}(t'') \right\rangle\right].
\end{eqnarray}
To evaluate the last term on the right hand side of Eq. \eqref{eq:tominimize} we again apply the Doob transform, obtaining
\begin{eqnarray}
\langle R^2\rangle &=& 2 t\langle \vec{c} \cdot \hat{D} \cdot \vec{c} \rangle + \\
&+&
2\int_0^t \D t' \int_{0}^{t'}\D t''\int\D \vec{x}~\D \vec{y}~ P(\vec{x},t'|\vec{y},t'')P^\st(\vec{y},t'') \left[\frac{\psi(\vec{x})}{P^{\st}(\vec{x})}  \frac{\psi(\vec{y})}{P^{\st}(\vec{y})} +  2 \frac{\psi(\vec{x})}{P^{\st}(\vec{x})} \vec{c}(\vec{y})  \cdot \hat{D}(\vec{y})\cdot \frac{\nabla_{\vec{y}} P(\vec{x},t'|\vec{y},t'') }{P(\vec{x},t'|\vec{y},t'')}\right]\nonumber
\end{eqnarray}
Integrating by parts the last term over $\vec{y}$ and substituting the definition of $\psi$ we obtain
\begin{eqnarray}\label{eq:R23}
\langle R^2\rangle &=& 2t \langle \vec{c} \cdot \hat{D} \cdot \vec{c} \rangle +\\
&+&2 \int_0^t \D t' \int_{0}^{t'} \D t'' \left\langle \frac{\vec{c}(\vec{x}) \cdot \vec{J}^{\st}(\vec{x}) + \vec{\nabla}_{\vec{x}} \cdot [\hat{D}(\vec{x}) \cdot \vec{c}(\vec{x}) P^{\st}(\vec{x})]}{P^{\st}(\vec{x})} \frac{\vec{c}(\vec{y}) \cdot \vec{J}^{\st}(\vec{y}) - \vec{\nabla}_{\vec{y}} \cdot [\hat{D}(\vec{y})\cdot \vec{c}(\vec{y}) P^{\st}(\vec{y})]}{P^{\st}(\vec{y})} \right\rangle \nonumber \\
\label{varRdev}
\end{eqnarray}
We introduce a new variable
\begin{equation}
\vec{\Gamma}(\vec{x}) = P^{\st}(\vec{x})\hat{D}(\vec{x}) \cdot \vec{c}(\vec{x}).
\end{equation}
 With this definition, the current is equal to the entropy production if $\vec{\Gamma}(\vec{x})=\vec{J}^{\st}(\vec{x})$. In terms
 of the new variable, the average of $R$ reads
\begin{equation}
\langle R \rangle = t \int \D\vec{x} ~\vec{\Gamma}(\vec{x}) \cdot \hat{D}^{-1} \cdot \frac{\vec{J}^{\st}(\vec{x})}{P^{\st}(\vec{x})}
\end{equation}
and, using Eq. \eqref{eq:R23}, the second moment can be expressed as
\begin{eqnarray}
\langle R^2\rangle &=& \int \D\vec{x} ~\D\vec{y} ~\left[ \vec{\Gamma}(\vec{x}) \cdot \hat{K}^{(1)} \cdot \vec{\Gamma}(\vec{y}) + \vec{\nabla}_{\vec{x}} \cdot \vec{\Gamma}(\vec{x}) K^{(2)} \vec{\nabla}_{\vec{y}} \cdot \vec{\Gamma}(\vec{y}) +\vec{\nabla}_{\vec{x}} \cdot \vec{\Gamma}(\vec{x}) \vec{K}^{(3)} \cdot \vec{\Gamma}(\vec{y}) 
+ \vec{\Gamma}(\vec{x}) \cdot \vec{K}^{(4)} \vec{\nabla}_{\vec{y}} \cdot \vec{\Gamma}(\vec{y}) \right] 
\label{R2K}
\end{eqnarray}
with
\begin{eqnarray}
\hat{K}^{(1)} &=& \frac{2 \hat{D}^{-1}(\vec{x})}{P^{\st}(x)} t \delta(\vec{x}-\vec{y}) \nonumber\\
&+& 2 \int_0^t \D t' \int_0^{t'} \D t'' P(x; t'|\vec{y}; t'') P^{\st}(\vec{y}) \frac{ \hat{D}^{-1}(\vec{x}) \cdot \vec{J}^{\st}(\vec{x})}{P^{\st}(\vec{x})^2} \frac{\vec{J}^{\st}(\vec{y}) \cdot \hat{D}^{-1}(\vec{y})}{P^{\st}(\vec{y})^2}  \\
K^{(2)} &=& - 2 \int_0^t \D t' \int_0^{t'} \D t'' P(x; t'|\vec{y}; t'') P^{\st}(\vec{y}) \frac{1}{P^{\st}(\vec{x}) P^{\st}(\vec{y})} \nonumber \\
\vec{K}^{(3)} &=& 2 \int_0^t \D t' \int_0^{t'} \D t'' P(x; t'|\vec{y}; t'') P^{\st}(\vec{y}) \frac{1}{P^{\st}(\vec{x})} \frac{\vec{J}^{\st}(\vec{y}) \cdot \hat{D}^{-1}(\vec{y}) }{P^{\st}(\vec{y})^2} \nonumber \\
\vec{K}^{(4)} &=& - 2 \int_0^t \D t' \int_0^{t'} \D t'' P(x; t'|\vec{y}; t'') P^{\st}(\vec{y}) \frac{\vec{J}^{\st}(\vec{x})\cdot \hat{D}^{-1}(\vec{x})}{P^{\st}(\vec{x})^2}  \frac{1}{P^{\st}(\vec{y})} \nonumber
\end{eqnarray}
Note that $\hat{K}^{(1)}$ is a tensor, $K^{(2)}$ is a scalar, and $\vec{K}^{(3)}$, $\vec{K}^{(4)}$ are vectors. We now compute the first variation respect to the function $\vec{\Gamma}(\vec{x})$
\begin{equation}\label{eq:firstvariation}
\delta \left( \frac{\langle R^2\rangle}{\langle R \rangle^2} \right) = \frac{1}{\langle R\rangle^2}\frac{\delta \langle R^2\rangle}{\delta \vec{\Gamma}}\delta \vec{\eta} - \frac{2 \langle R^2\rangle}{\langle R\rangle^3}\frac{\delta \langle R\rangle}{\delta \vec{\Gamma}}\delta \vec{\eta}
\end{equation}
where
\begin{equation}\label{eq:avr}
\frac{\delta \langle R\rangle}
{\delta \vec{\Gamma}}\delta \vec{\eta} = t \int \D\vec{x}~ \vec{\eta}(\vec{x})\cdot \hat{D}^{-1} \cdot \frac{\vec{J}^{\st}(\vec{x})}{P^{\st}(\vec{x})}
\end{equation}
while
\begin{eqnarray}
\frac{\delta \langle R^2\rangle}
{\delta \vec{\Gamma}}\delta \vec{\eta} &=&\int \D\vec{x} \D\vec{y} ~ \Bigg[  \vec{\eta}(\vec{x}) \cdot \hat{K}^{(1)} \cdot \vec{\Gamma}(\vec{y}) +  \vec{\Gamma}(\vec{x}) \cdot \hat{K}^{(1)} \cdot \vec{\eta}(\vec{y})  \nonumber\\
&-&\vec{\eta}(\vec{x})\cdot\left(\vec{\nabla}_{\vec{x}} K^{(2)}\right) \vec{\nabla}_{\vec{y}}\cdot\vec{\Gamma}(\vec{y})-\vec{\nabla}_{\vec{x}}\cdot\vec{\Gamma}(\vec{x})\left(\vec{\nabla}_{\vec{y}} K^{(2)}\right)\cdot\vec{\eta}(\vec{y}) \nonumber\\
&-&\vec{\eta}(\vec{x}) \cdot \vec{\Gamma}(\vec{y}) \left(\vec{\nabla}_{\vec{x}} \cdot \vec{K}^{(3)}\right)+\vec{\nabla}_{\vec{x}} \cdot \vec{\Gamma}(\vec{x})\vec{K}^{(3)} \cdot \vec{\eta}(\vec{y})\nonumber\\
&+& \vec{\eta}(\vec{x}) \cdot \vec{K}^{(4)} \vec{\nabla}_{\vec{y}} \cdot \vec{\Gamma}(\vec{y}) - \vec{\Gamma}(\vec{x}) \cdot \vec{\eta}(\vec{y}) \left(\vec{\nabla}_{\vec{y}} \cdot \vec{K}^{(4)}\right)\Bigg]
\end{eqnarray}
where we already integrated by parts.
We reorganize this expression  by swapping the variables $x$ and
$y$ in the appropriate terms and noting that $K^{(1)}$ and
$K^{(2)}$ are self-adjoint, whereas $K^{(3)T}=-K^{(4)}$. Here the superscript ``$T$'' denote the transposed operator, i.e., the operator obtained by swapping $x$ and $y$. This results in
\begin{eqnarray}\label{eq:avr2}
\frac{\delta \langle R^2\rangle}
{\delta \vec{\Gamma}}\cdot\delta \vec{\eta} &=& \int \D\vec{x}~ \vec{\eta}(\vec{x}) \cdot \vec{G}(\vec{x},t) \nonumber \\
\end{eqnarray}
with
\begin{equation}
\vec{G}(\vec{x},t) =2 \int \D\vec{y} ~ \hat{K}^{(1)} \cdot \vec{\Gamma}(\vec{y})-\left(\vec{\nabla}_{\vec{x}} K^{(2)}\right) \vec{\nabla}_{\vec{y}} \cdot \vec{\Gamma}(\vec{y}).
\end{equation}
Imposing from Eqs.~\eqref{eq:firstvariation} that the first variation vanishes for any choice of $\eta$ and using  Eqs. \eqref{eq:avr}, and \eqref{eq:avr2} leads to the condition
\begin{equation}
\vec{G}(\vec{x},t) = 2 t \hat{D}^{-1}(\vec{x}) \cdot \frac{\vec{J}^{\st}(\vec{x})}{P^{\st}(\vec{x})} \frac{\langle R^2\rangle}{\langle R \rangle}
\label{firstvariationnull}
\end{equation}

Substituting the expression of $\vec{G}(\vec{x},t)$ and expressing the equation in terms of $\vec{c}_\hyp(\vec{x})$ we finally obtain Eq.~(13) in the Main Text.

\section{Long-time limit}
\label{IV}

We now derive the long-time limit of Eq.~(13) in the Main Text, that we rewrite as
\begin{eqnarray}\label{eq:ltlSI}
P^{\st}(\vec{x}) \hat{D}(\vec{x})\cdot\vec{c_\hyp}(\vec{x}) &+& 
 \frac{P^{\st}(\vec{x})}{t}\hat{D}(\vec{x})\cdot \vec{\nabla}_{\vec{x}} \left\{ \int \D\vec{y} \int_0^t \D t' \int_0^{t'} \D t'' ~\frac{P(\vec{x}; t'|\vec{y}; t'')}{P^{\st}(\vec{x})} \vec{\nabla}_{\vec{y}} \cdot \left[ P^{\st}(\vec{y}) \hat{D}(\vec{y}) \cdot \vec{c}_\hyp(\vec{y}) \right] \right\} +\\
&+& \frac{\vec{J}^{\st}(\vec{x})}{t~P^{\st}(\vec{x})} \int \D\vec{y} \int_0^t \D t' \int_0^{t'} \D t'' P(\vec{x}; t'|\vec{y};t'') \vec{J}^{\st}(\vec{y}) \vec{c}_\hyp(\vec{y}) 
=\vec{J}^{\st}(\vec{x}) \frac{\langle R_\hyp^2 \rangle}{2 \langle R_\hyp \rangle}\nonumber
\end{eqnarray}
In the limit $t\rightarrow\infty$, the first two term on the left-hand side of Eq.~\eqref{eq:ltlSI} converge to a finite value, whereas the last term on the left-hand side and the right-hand side diverge as $t$. 
To avoid this divergence, we subtract from both sides the contribution $\vec{J}^{\st}(\vec{x})\langle R_\hyp\rangle/2$, that we rewrite as
\begin{equation}
\frac{\vec{J}^{\st}(\vec{x}) \langle R_\hyp \rangle}{2} = \frac{\vec{J}^{\st}(\vec{x})}{t} \int_0^t \D t' \int_0^{t'} \D t'' \left\langle \frac{\vec{c}_\hyp \cdot \vec{J}^{\st} }{P^\st}\right\rangle = \frac{\vec{J}^{\st}(\vec{x})}{t~P^{\st}(\vec{x})} \int_0^t \D t' \int_0^{t'} \D t'' \int \D \vec{y}~ P^{\st}(\vec{x}) \vec{c}_\hyp(\vec{y}) \cdot \vec{J}^{\st}(\vec{y})
 \end{equation}
Equation~\eqref{eq:ltlSI} then becomes
\begin{eqnarray}
P^{\st}(\vec{x}) \hat{D}(\vec{x})\cdot\vec{c_\hyp}(\vec{x})  &+& 
\frac{P^{\st}(\vec{x})}{t}\hat{D}(\vec{x})\cdot \vec{\nabla}_{\vec{x}} \left\{ \int \D\vec{y} \int_0^t \D t' \int_0^{t'} \D t'' ~\frac{P(\vec{x}; t'|\vec{y}; t'')}{P^{\st}(\vec{x})} \vec{\nabla}_{\vec{y}} \cdot \left[ P^{\st}(\vec{y}) \hat{D}(\vec{y}) \cdot \vec{c}_\hyp(\vec{y}) \right] \right\} +\\
&+& \frac{\vec{J}^{\st}(\vec{x})}{t P^{\st}(\vec{x})} \int \D\vec{y} \int_0^t \D t' \int_0^{t'} \D t'' \left[ P(\vec{x}; t'|\vec{y}; t'') - P^{\st}(\vec{x}) \right] \vec{J}^{\st}(\vec{y}) \cdot \vec{c}_\hyp(\vec{y})  = \vec{J}^{\st}(\vec{x}) \frac{\sigma^2_{R_\hyp}}{2 \langle R_\hyp \rangle}.\nonumber
\end{eqnarray}
At this point it is safe to take the limit $t\rightarrow \infty$. Performing the integral over time and using that at steady state the propagator depends only on the time difference $t'-t''$ we obtain
\begin{eqnarray}
P^{\st}(\vec{x}) \hat{D}(\vec{x})\cdot\vec{c_\hyp}(\vec{x}) &+& P^{\st}(\vec{x}) \left\{ \int \D\vec{y} ~ \hat{D}(\vec{x})\cdot\vec{\nabla}_{\vec{x}}\left[\frac{\phi(\vec{x},\vec{y})}{P^{\st}(\vec{x})} \right]\vec{\nabla}_{\vec{y}} \cdot \left[ P^{\st}(\vec{y}) \hat{D}(\vec{y}) \cdot \vec{c}_\hyp(\vec{y}) \right] \right\}+\\
&+&\frac{\vec{J}^{\st}(\vec{x})}{P^{\st}(\vec{x})} \int \D\vec{y} ~\phi(\vec{x},\vec{y}) \vec{J}^{\st}(\vec{y}) \cdot \vec{c}_\hyp(\vec{y})  = \vec{J}^{\st}(\vec{x}) \frac{\sigma^2_{R_\hyp}}{2 \langle R_\hyp \rangle}
\label{eq:ltlFIN}
\end{eqnarray}
where the function $\phi(\vec{x},\vec{y})$ is defined in Eq.~(16) of the Main Text. Integrating the second term on the right hand side of Eq.~\eqref{eq:ltlFIN} by parts and rewriting the whole expression as an integral equation leads to Eqs.~(14) and (15) of the Main Text.

\section{Discretization of the integral kernel in one dimension}
\label{V}

In this Section we discretize the integral kernel of the one dimensional model in the Main Text and show how to solve the Euler-Lagrange equation numerically in this case. We start by writing the expression of kernel in Eq.~(17) in one dimension
\begin{eqnarray}
K(x,y) = \frac{(J^{\st})^2}{P^{\st}(x)} \phi(x,y) + P^{\st}(x) D(x) \delta(x-y) - D(x) P^{\st}(x) \nabla_{xy} \left[ \frac{\phi(x,y)}{P^{\st}(x)} \right] P^{\st}(y) D(y)
\label{LTL}
\end{eqnarray}
where we used that the the stationary flux is constant in one dimension. 

Writing explicitly the spatial derivatives results in
\begin{eqnarray}
K(x,y) = \frac{(J^{\st})^2}{P^{\st}(x)} \phi(x,y) + P^{\st}(x) D(x) \delta(x-y) + \left[ - \nabla_{xy} \phi(x,y) + \frac{\nabla_x P^{\st}(x)}{P^{\st}(x)} \nabla_y \phi(x,y) \right] D(x) P^{\st}(y) D(y)
\label{LTL2}
\end{eqnarray}

The system is periodic in the interval $[0,1]$. We discretize this interval by introducing a mesh $\Delta$. In this way, a function of $x$ becomes a function of a discrete index $i=\lfloor x/\Delta \rfloor$, where $\lfloor\dots\rfloor$ denotes the integer part. Similarly, functions of $x$ and $y$ become matrices with indices $i$ and $j$. In particular, the integral Eq. (14) in the Main Text becomes a linear system
\begin{equation}
\sum_j M_{ij}(c_\hyp)_j=J^\st
\end{equation}
where we call $M_{ij}$ the matrix obtained by discretizing the kernel of the integral equation. Such matrix reads
\begin{eqnarray}
M_{ij}  = \frac{(J^{\st})^2}{P^{\st}_i} (\phi)_{ij} + P^{\st}_i D_i \delta_{ij} + \left( - (\nabla_{xy} \phi)_{ij} + \frac{(\nabla_x P^{\st})_i}{P^{\st}_i} (\nabla_y \phi)_{ij} \right) D_i P^{\st}_j D_j
\label{LTL2b}
\end{eqnarray}
where $\delta_{ij}$ is the Kronecker delta. We use the notation $(\nabla_x P^{\st})_i$, $(\nabla_{y} \phi)_{ij}$, and $(\nabla_{xy} \phi)_{ij}$ for the discretized derivatives, that are defined as
\begin{eqnarray}
(\nabla_x P^{\st})_{i} &=& \frac{-P^{\st}_{i+2} + 8 P^{\st}_{i+1} - 8 P^{\st}_{i-1} + P^{\st}_{i-2}}{12 \Delta} + \mathcal{O}(\Delta^4) \nonumber \\
(\nabla_y \phi)_{ij} &=& \frac{-\phi_{i,j+2} + 8 \phi_{i,j+1} - 8 \phi_{i,j-1} + \phi_{i,j-2}}{12 \Delta} + \mathcal{O}(\Delta^4) \nonumber \\
(\nabla_{xy} \phi)_{ij} &=& \frac{-(\nabla_y \phi)_{i+2,j} + 8 (\nabla_y \phi)_{i+1,j} - 8 (\nabla_y \phi)_{i-1,j} + (\nabla_y \phi)_{i-2,j}}{12 \Delta} + \mathcal{O}(\Delta^4)
\end{eqnarray}
In the $1D$ example presented in the Main Text, we set $D(x) = 1$. To estimate the propagator $P(x;t|y;0)$, we numerically solve the Fokker-Planck equation associated to Eq.~(21) of the Main Text 
\begin{equation}
\partial_t P(x;t|y;0) = - \nabla_x \left[ (\nabla_x (f x - U(x)) P(x;t|y;0) \right] + \nabla_x^2 P(x;t|y;0).
\end{equation}
For the numerical integration, we approximate the initial condition $P(x;0|y;0) = \delta(x-y)$ with a Gaussian distribution with mean $y$ and variance $3 \times 10^{-6}$. We use the built-in solver in Mathematica with a spatial mesh $\Delta = 0.002$, a time step $\Delta t = 10^{-5}$ and an ``accuracy goal" equal to half the Machine Precision (53 bits). We reach stationarity (with an error on the order of $10^{-5}$) after about $10^5$ time steps, i.e., at a final time $t_f=1$. We compute $\phi(x,y)$ from the propagator by integrating over time using the trapezoidal rule. Stationary flux and probability distribution are computed from the stationary solution $P(x;t_f|y;0) = P^\st(\vec{x}) + \mathcal{O}(10^{-5})$.

\section{Hyperaccurate current for the two dimensional model}
\label{VI}

In this Section we derive an explicit expression for the hyperaccurate current for the two-dimensional model in Eq.~(24) of the Main Text. First of all, $\vec{c}_\hyp$ can not depend on $x_1$ because of invariance under translations along the $x_1$ axis. The Euler-Lagrange equations for this model then read
\begin{gather}
\label{tosolve}
c_{\hyp,1}(x_2) + f^2 \int \D y_2 \cos(2 \pi x_2) \phi(x_2,y_2) \cos(2 \pi y_2) c_{\hyp,1}(y_2) = f \cos(2 \pi x_2) \\
\label{2component}
c_{\hyp,2}(x_2) = - \int \D y_2 ~\partial_{x_2,y_2} \phi(x_2, y_2) ~c_{\hyp,2}(y_2)
\end{gather}
where $\phi(x_2,y_2)$ can be computed explicitly from the diffusion equation 
\begin{equation}
\phi(x_2,y_2) = \sum_{n=0}^{+\infty} \frac{1}{2 \pi^2 n^2} \cos(2 \pi n (x_2 - y_2))
\end{equation}
Eq.~\eqref{2component} is solved by  $c_{\hyp,2}(x_2) = 0$, so that the hyperaccurate vector field $\vec{c}_\hyp(x_2) = (c_{\hyp,1}(x_2),0)$ is governed by the one-dimensional Eq.~\eqref{tosolve}.
We now expand the solution in a Fourier basis
\begin{equation}
c_{\hyp,1}(y) = \kappa_{\hyp,0} + \chi_{\hyp,0} + \sum_{n=1}^{+\infty} \kappa_{\hyp,n} \cos(2 \pi n y) + \sum_{n=1}^{+\infty} \chi_{\hyp,n} \sin(2 \pi n y)
\label{solution}
\end{equation}
Inserting Eq.~\eqref{solution} in Eq.~\eqref{tosolve}, we get
\begin{gather}
\kappa_{\hyp,0} + \chi_{\hyp,0} + \sum_{n=1}^{+\infty} \kappa_{\hyp,n} \cos(2 \pi n y_2) + f^2 \cos(2 \pi y_2) \sum_{n=2}^{+\infty} \frac{1}{2 \pi^2 n^2} \frac{\kappa_{\hyp,n+1}+\kappa_{\hyp,n-1}}{4} \cos(2 \pi n y_2) + \frac{f^2}{2 \pi^2} \cos(2 \pi y_2)^2 \left( \frac{\kappa_{\hyp,2}}{4} + \frac{\kappa_{\hyp,0}}{2} \right) + \nonumber \\
+ \sum_{n=1}^{+\infty} \chi_{\hyp,n} \sin(2 \pi n y_2) + f^2 \cos(2 \pi y_2) \sum_{n=2}^{+\infty} \frac{1}{2 \pi^2 n^2} \frac{\chi_{\hyp,n+1}+\chi_{\hyp,n-1}}{4} \sin(2 \pi n y_2) + \frac{f^2}{2 \pi^2} \frac{\chi_{\hyp,2}}{4} \cos(2 \pi y_2) \sin(2 \pi y_2) = f \cos(2 \pi y_2)
\end{gather}
To make progress, we use the properties  of trigonometric functions
\begin{eqnarray}
\int_0^1 \D y_2 \cos(2 \pi n (x_2-y_2)) \cos(2 \pi y_2) \cos(2 \pi m y_2) &=& \frac{1}{4} \cos(2 \pi n y) \delta_{m,n\pm 1} + \frac{1}{4} \cos(2 \pi y) \delta_{n,1} \delta_{m,0} \\
\int_0^1 \D y_2 \cos(2 \pi n (x_2-y_2)) \cos(2 \pi y_2) \sin(2 \pi m y_2) &=& \frac{1}{4} \sin(2 \pi n y) \delta_{m,n\pm 1} - \frac{1}{4} \sin(2 \pi y) \delta_{n,1} \delta_{m,0}\\
\cos(2 \pi n y_2)\cos(2 \pi y_2) &=& \frac{1}{2} \bigg(\cos(2 \pi (n+1) y_2) + \cos(2 \pi (n-1) y_2) \bigg) \\
\sin(2 \pi n y_2)\cos(2 \pi y_2) &=& \frac{1}{2} \bigg(\sin(2 \pi (n+1) y_2) + \sin(2 \pi (n-1) y_2) \bigg)
\end{eqnarray}
Expressing the products of trigonometric functions using these relations we obtain
\begin{gather}
\kappa_{\hyp,0} + \chi_{\hyp,0} + \sum_{n=1}^{+\infty} \kappa_{\hyp,n} \cos(2 \pi n y_2) + f^2 \sum_{n=2}^{+\infty} \frac{\kappa_{\hyp,n+1}+\kappa_{\hyp,n-1}}{16 \pi^2 n^2} \bigg(\cos(2 \pi (n+1) y_2) + \cos(2 \pi (n-1) y_2) \bigg) + \nonumber \\ + \frac{f^2}{4 \pi^2} \bigg(\cos(4 \pi y_2) + 1 \bigg) \left( \frac{\kappa_{\hyp,2}}{4} + \frac{\kappa_{\hyp,0}}{2} \right) + \nonumber \\
+ \sum_{n=1}^{+\infty} \chi_{\hyp,n} \sin(2 \pi n y_2) + f^2 \sum_{n=2}^{+\infty} \frac{\chi_{\hyp,n+1}+\chi_{\hyp,n-1}}{16 \pi^2 n^2} \bigg(\sin(2 \pi (n+1) y_2) + \sin(2 \pi (n-1) y_2) \bigg) + \nonumber \\
+ \frac{f^2}{4 \pi^2} \frac{\chi_{\hyp,2}}{4} \bigg(\sin(4 \pi y_2) + 1 \bigg) = f \cos(2 \pi y_2)
\end{gather}
The coefficients associated to the sine vanish, $\chi_m = 0$, since on the r.h.s we have the stationary flux, which is a cosine. The non-vanishing coefficients can be written at any order
\begin{eqnarray}
\kappa_{\hyp,0} &+& \frac{f^2}{4 \pi^2} \left( \frac{\kappa_{\hyp,2}}{4} + \frac{\kappa_{\hyp,0}}{2} \right) = 0 \nonumber \\
\kappa_{\hyp,1} &+& \frac{f^2}{8 \pi^2} \frac{\kappa_{\hyp,3} + \kappa_{\hyp,1}}{8} = f \nonumber \\
\kappa_{\hyp,2} &+& \frac{f^2}{18 \pi^2} \frac{\kappa_{\hyp,4}+\kappa_{\hyp,2}}{8} + \frac{f^2}{4 \pi^2} \left( \frac{\kappa_{\hyp,2}}{4} + \frac{\kappa_{\hyp,0}}{2} \right) = 0 \nonumber \\
\kappa_{\hyp,n} &+& f^2 \frac{\kappa_{\hyp,n} + \kappa_{\hyp,n-2}}{16 \pi^2 (n-1)^2} + f^2 \frac{\kappa_{\hyp,n+2} + \kappa_{\hyp,n}}{16 \pi^2 (n+1)^2} = 0 \;\;\;\;\ \mbox{for } n > 2
\label{setofeq}
\end{eqnarray}
where we split the second summation and changed the indices in $m = n+1$ and $m = n-1$.
The curves in Fig.2b of the Main Text were obtained by truncating the expansion to the $30$th order.  Higher order coefficients were smaller than $10^{-9}$ for all values of the force in the explored range $f\in[0.1,100]$.

\end{widetext}

\end{document}